**Understanding the PULSAR Effect in Combined Radiotherapy and Immunotherapy through Attention Mechanisms with a Transformer Model**

Hao Peng[1,2], Casey Moore[1], Debabrata Saha[1], Steve Jiang[1,2] and Robert Timmerman[1]

[1]Departments of Radiation Oncology, University of Texas Southwestern Medical Center, Dallas, Texas

[2] Medical Artificial Intelligence and Automation Laboratory, University of Texas Southwestern Medical Center, Dallas, Texas

Email: hao.peng@utsouthwestern.edu

**Abstract**

PULSAR (personalized, ultra-fractionated stereotactic adaptive radiotherapy) is the adaptation of stereotactic ablative radiotherapy towards personalized cancer management. For the first time, we applied a transformer-based attention mechanism to investigate the underlying interactions between combined PULSAR and PD-L1 blockade immunotherapy based on a murine cancer model (Lewis Lung Carcinoma, LLC). The proposed approach is able to predict the trend of tumor volume change semi-quantitatively, and excels in identifying the potential causal relationships through both self-attention and cross-attention scores.

**Introduction**

The field of combining radiotherapy and immunotherapy is rapidly evolving, and one aspect of particular interest is determination of optimal timing and sequence to harness the potential synergy between radiation therapy and immune checkpoint blockade. Despite promising progress, combination of the two treatments has yielded lackluster results in both clinical and preclinical studies[1-10]. Defining the ideal combination and synergy continues to pose a considerable challenge. Our team at UT Southwestern Medical Center (UTSW) is exploring PULSAR (personalized, ultra-fractionated stereotactic adaptive radiotherapy), which aims to deliver tumoricidal doses in a pulsed mode with long intervals[11]. Longer intervals spanning weeks or months not only allow for greater recovery of normal tissue following an injury, but may also maximize potential synergies resulting from concomitant immune-oncology approaches. As adaptive immune response typically takes considerable time to develop and reach its full effectiveness, exploring its temporal interaction with pulsed, more independent radiation doses (henceforth referred to as the PULSAR effect), presents an intriguing opportunity.

One aspect of interest is determining the optimal timing and sequence to harness the immune-mediated antitumor effects resulting from radiation therapy. For instance, one clinical trial studied the outcome of multisite, more conventionally delivered Stereotactic Body Radiation Therapy (SBRT) followed by pembrolizumab, reporting an overall response rate of 13.2% in advanced solid tumors[6]. Another PEMBRO-RT phase 2 clinical trial in non-small cell lung cancer demonstrated a doubling in overall response when patients were treated with SBRT (3×8 Gy) followed by pembrolizumab[7]. The synergistic impact has been extensively explored in preclinical models as well. One study delivered 10 to 24 Gy in 1 to 3 daily, or every other day fractions and began PD1/PD-L1 checkpoint blockade therapy within a day of radiation[5]. Another preclinical study included 3 daily SBRT fractions of 8 Gy, followed by anti-CTLA4 treatment, beginning on the day of the last fraction[8]. Both studies demonstrated additive benefits when immune checkpoint inhibitor (ICI) blockade is given concomitantly with radiation. Moreover, there are conflicting reports regarding the optimal timing of PD-L1 therapy in relation to radiation. While some studies showed no additive benefit when PD-L1 therapy was administered 6 days after a single dose of radiation[9], others suggested clear benefits of PD-L1 therapy given every 21 days after radiation[10].

Given the complexities of both physical and biological processes involved, it is an incredibly challenging task to model the temporal synergy between two treatments. Instead of going with a conventional analytical approach (e.g., solving differential questions or linear equations), we proposed an AI approach based on a transformer model with attention mechanisms. We adopted this concept from neural machine translation, where a model translates a sentence from a source language to a sentence in a target language[12, 13]. In this study, each *"word in a sentence"* corresponds to an input, representing either a radiation pulse or an anti-PD-L1 dose (**Fig. 1a**). The radiation and administration of anti-PD-L1 are viewed as two external stimulation signals occurring in a temporal sequence. We developed the transformer model to accomplish two objectives: 1) predict tumor changes for a given treatment scheme, and 2) provide self-attention and cross-attention maps. We hypothesize that the cross-attention maps in our study are the biological equivalent representation of the semantic similarity between source and target sentences in neural translation. To a certain extent, the second objective holds greater significance than the first, as it would offer more insights regarding the causal relationships of the PULSAR effect.

**Methods and Materials**

**Problem Formulation**

Given the complexities of interactions involved in radioimmunotherapy, it is challenging to model the response of individual components such as tumor growth, radiation damage, T-cell infiltration, and signaling pathways. To understand this better, several characteristics of involved physical and biological processes are summarized below. For instance, tumor growth with neither radiation nor PD-L1 inhibitor follows an exponential model. The killing effect of a single radiation pulse is commonly modeled through a linear-quadratic (LQ) model or a universal survival model[14]. After the radiation is delivered on a given day, the repopulation of tumor cells is time dependent and also follows an exponential model, dependent on the half-time of repopulation (from less than a day to several weeks) and lag time[15]. The interplay between radiation and checkpoint inhibitors such as PD-L1 antibody is even more complicated. On one hand, high dose radiation promotes both local and systemic immune responses and recruits CD8+ cytotoxic T cells to tumor sites by mechanism, such as the c-GAS-STING cytosolic DNA-sensing pathway[16]. The recruitment process is time-dependent and takes up to several days. On the other hand, interferon gamma released by CD8+ T cells, leads to compensatory PD-L1 expression on tumor as a mechanism to prevent autoimmunity. Meanwhile, $T_{reg}$ cells inhibit immune response by elevating the expression of CTLA4 to inhibit the activities of antigen-presenting cells, and releasing cytokines such as TGF-β and IL10 to suppress the functions of effector T cells[17-19]. As a result, the net tumoricidal effect of combined therapy, depends on radiation dose, anti-PD-L1 dose, time, T cell dynamics and tumor microenvironment. Furthermore, as radiation pulses are applied and T cells are continuously recruited to tumor sites, radiations kill both tumor and T cells. However, there might be a difference between tumor-resident T cells and newly arrived T cells from lymph nodes in terms of radiosensitivity[20].

**Small Animal Experiments**

In this study, the impact of radiation schedules on tumor growth within an immune-resistant mouse model was studied (**Fig. 1b**). Recognized as a "cold" tumor with low infiltration of T cells but high infiltration of myeloid-derived suppressor cells (MDSCs), Lewis lung carcinoma (LLC) offers valuable insights into the temporal behaviors of the adaptive immune response, as it takes a longer time to develop and reach its full effectiveness[21]. In addition, differing from most preclinical models using radiation and with either daily or every other day fractions[5, 8-10], we tested radiation pulses with a longer spacing. The experimental protocols in this manuscript were approved by the Institutional Animal Care and Use Committee (IACUC) at UT Southwestern Medical Center (UTSW). The authors confirm that all animal procedures were performed in accordance with the animal experimental guidelines set by the IACUC at UTSW. UTSW uses the "Guide for the Care and Use of Laboratory Animals" when establishing animal research standards. The authors also confirm that the study was reported in accordance with ARRIVE guidelines (https://arriveguidelines.org). More details about the PULSAR study design can be found in Moore et al[11]. C57BL/6J mice were used, and LLC was derived from lung cancer of the C57BL/6 line. Tumor cells were injected subcutaneously on the right leg of mice. 10F.9G2 was used as an isotype control for anti-PD-L1. Mice were administered (i.p) 200 μg α-PD-L1 or 200 μg isotype control according to different schedules (**Table S1**). Local irradiations were conducted on a dedicated x-ray irradiator (X-RAD 32, Precision X-ray, Inc.) (Supplemental Materials). The tumor volumes were measured by length (x), width (y), and height (z), and calculated as tumor volume = xyz/2. If each length, width or height of tumor was larger than 2 cm, the tumor volume was larger than 1500 mm$^3$, or the mouse had significant ulceration in the tumor (see supporting materials), this indicated that the mouse reached the survival endpoint and so was euthanized by exposure to $CO_2$. As a result, the data collection and volume measurement towards the end of the study showed an increased degree of variance, which was one of our study's limitations. As noted in the study by Moore et al.[11], the majority of experiments demonstrated no improvement in growth control with the addition of anti-PD-L1 therapy compared to controls. However, certain combinations of pulsed radiation with anti-PD-L1 did slow growth. This somewhat unexpected synergy between radiation and ICI constitutes the PULSAR effect.

**Data Pre-processing**

For each treatment group, seven or eight animals were studied, and the measurement of tumor volume exhibited large variance. From a machine learning perspective, this weakens the one-to-one relationship between an input and output sequence. Consequently, data augmentation was necessary for training the model. The details of data processing and feature extraction are summarized below. Although the measurement was performed up to 40 days for some animals, an increased percentage of animals had missing data towards the survival endpoint due to severe ulceration. Therefore, the total time course was set to be 28 days (four weeks) for all samples to ensure data consistency. There were 26 treatment groups in total. For each group, the mean and standard deviation were first calculated. Then 50 samples were randomly generated, assuming the average volume following a Gaussian distribution. To address the issue of large variance, the standard deviation of each group was intentionally reduced by a factor of five, which we believe was achievable by employing more accurate volume contouring techniques and increasing the number of animal samples per group. To a certain extent, the generation of 50 samples through data augmentation can be regarded as a strategy to help avoid overfitting and reinforce good generalization. A total of 1300 sequences were utilized for training (50 samples per group, 26 groups), and each sequence consisted of 28 steps (days), with two inputs and one output. The two inputs corresponded to radiation and anti-PD-L1, respectively. The output sequence represented the tumor volume change, and if no input or output was available at a step, the values were set to zero.

**Transformer Model with Attention**

The concept was adopted from machine translation, where a model translates a sentence from a source language to a sentence in a target language. In simpler terms, the transformer model focuses attention on specific input words, and the attention acts as a link connecting the encoder and decoder. Each word in an input sentence is assigned its own query, key, and value vectors. These vectors are generated by multiplying the encoder's representation of the particular word with three distinct weight matrices developed during training. In this study, each word in a sentence corresponds to an input, representing either a radiation pulse or an anti-PD-L1 dose. The execution of an attention layer involves five steps: 1) preparing hidden states and deriving a score for the encoder hidden state, 2) passing all scores through a softmax layer to generate attention distribution, 3) multiplying each encoder hidden state by its softmax score to acquire the alignment vector or annotation vector, 4) summing the alignment vectors to obtain the aggregated information from the previous step, and 5) inputting the context vector into the decoder. Due to the small size of the dataset, the transformer model was kept at a very basic form with only 536 parameters. Both self-attention in the encoder (radiation) and cross-attention between the encoder and decoder were investigated. The latter helped in studying the relationship between radiation and immunotherapy concerning elapsed days. Given the small size of the dataset, the transformer model was kept in its basic form (**Fig. 1**). Every *"word"* represented a five-bit vector, with one bit indicating the physical input (based on the presence of anti-PD-L1 or radiation dose) and four bits for positional embedding. A causal mask was applied to prevent the attention mechanism from sharing information about tokens at future positions ("don't look ahead"), enforcing the autoregressive property during both training and inference. The decoder was constructed similarly, differing only in the output stage where a full connection was employed to predict tumor volume. We developed the transformer model using the Keras package.

**Model Training**

The loss function was calculated based on the non-zero points (e.g. with measurements performed) only. The AdamW algorithm with default settings was used for optimization (learning rate = 0.0001, weight decay = 0.0001, beta1 = 0.9, beta2 = 0.999, epsilon = $1\times10^{-8}$)[22]. The batch size was 32 and the total epoch iteration was 5000. In total, 1200 samples (24 groups, 50 samples in each group) were generated from model training (80%) and testing (20%). To ease the learning task, the volume change between two adjacent measurement points was used as output. The L2 loss was used, representing the discrepancy between the predicted volume change (VC) (represented by Y in **Eq.1**) and experimentally measured VC. The VC difference (ΔVC) assumed either positive or negative values. If tumor volume was measured on 6 different days, we would have 5 ΔVC values. To assess the generalization capability and mitigate potential overfitting, holdout cross-validation was conducted. This involved testing the transformer model on 64 randomly sampled samples that were not included in the training dataset.

$$VC_i = Volume_{i+1} - Volume_i \qquad \Delta VC_i = Y_i - VC_i \qquad (1)$$

## Results

### Prediction of Tumor Volume Change

Tumor volumes measured at multiple time points served as the dataset for training the transformer model. In **Figure 2**, the prediction of tumor volume change is presented for groups 1 through 12. See the results for groups 13 through 26 in supporting materials. Each bar represents the change between two adjacent measurement time points, and combining values from five bars would provide the overall volume change for a specific treatment. The training and validation loss converged after around 5000 epochs, indicating no noticeable overfitting. For all 26 groups, good agreement between predicted and experimental outcomes is observed. The discrepancy of a single bar falls within the range of -68 cc to 61 cc. Considering the tumor volume reaching approximately 600 cc by day 20 without any treatment, the transformer model demonstrates satisfactory predictive accuracy. This positions it as a valuable in-silico modeling tool for simulating diverse sequences not included in the experiments.

### Self-attention Maps

**Figure 3a** presents the self-attention maps (radiation vs. radiation). For both group 6 (40Gyd1, a single radiation of 40 Gy on day 1) and group 4 (20Gyd1), the impact of radiation is observable and persists for an extended period of up to 20 days, with the former exhibiting a higher attention score due to a higher dose. The comparison between group 4 and group 8 (20Gyd1+20Gyd10) shows that the second radiation pulse shifts the attention by 10 days and mitigates the attention score of the first radiation pulse after day 10. Similarly, between group 10 (10Gyd1+10Gyd2) and group 12 (10Gyd1+10Gyd10), the attention is shifted by 10 days when the two radiation pulses are spaced by 10 days. The cause of such an attention shift is not clear, and may be simply attributed to the direct killing of both tumor cells and T cells due to the second radiation, *"resetting"* the system to a new state.

**Cross-attention Maps**

The cross-attention plots in **Figure 3b** provide mechanistic interpretations of the evolutional dynamics of the interaction between two treatments (radiation vs. anti-PD-L1). Despite the absence of direct biological meaning, they may still reveal whether an interaction is strong or weak, how long an interaction develops, as well how an interaction changes when a radiation pulse is applied at a specific time point. To gain further insights into the respective contributions, the difference map between the two groups pinpoints the sole contribution of either radiation or anti-PD-L1. For instance, both group 1 and group 3 were administered the isotype control (no anti-PD-L1), but only group 3 experienced a single 20 Gy pulse on day 1. The difference in the attention score (Group 3- Group 1) is thus solely attributed to radiation, which peaks on day 4 and diminishes gradually over time, as evidenced by the trend in the fourth row. It is important to note that this relationship is not one-to-one, but rather one-to-many, signifying that an output is influenced by more than one input stimulus, each with varying degrees of importance. Another interesting pattern arises in the comparison between group 4 (matching group 3, but with anti-PD-L) and group 2 (matching group 1, but with anti-PD-L1). The impact of concurrent radiation with anti-PD-L1 is found to peak at day 10, implying the occurrence of maximum synergy. Such a pattern also emerges in the comparison between group 4 and group 3, where the effect is ascribed solely to anti-PD-L1 with the presence of a single radiation dose at the onset. These results serve as evidence that the PULSAR effect is not instantaneous but takes time to build up.

Three additional examples, all with the presence of anti-PD-L1, further demonstrate the dependency of PULSAR effect on radiation dose and scheduling. Between group 6 (40 Gyd1) and group 4 (20 Gyd1), the previous one displays higher attention (positive difference) in most pixels, except for a few pixels where there is a negative difference. This suggests that although a dose of 40 Gy has a more pronounced killing effect, it may not necessarily lead to optimal synergy due to potential overkilling of surrounding T cells when compared to a 20 Gy dose. In the comparison between group 6 and group 8 (20Gyd1+20Gyd10), the left half of the attention map exhibits the same pattern as in Group 6 - Group 4 (note the slightly different scale of the color bar). One noticeable pattern emerges in the region corresponding to the second 20 Gy pulse (10th row): the attention flips its sign for multiple pixels, indicative of the PULSAR effect triggered by the second pulse. This suggests that when the single pulse of 40 Gy is divided into two pulses of 20 Gy each, the PULSAR effect changes. Another example is shown in the comparison between group 16 (10Gyd1+10Gyd2) and group 12 (10Gyd1+10Gyd10), illustrating the dependence of the PULSAR effect on the spacing (1 day vs. 10 days).

**Discussion**

Modeling the interaction and synergy between pulsed radiation and ICI therapy is a challenging task. For the first time, we attempted to tackle that challenge employing an AI approach. The transformer model, coupled with the attention mechanism, leverages its unique capacity to process sequential information and capture underlying correlations[12, 13]. As illustrated in **Figure 2**, it is able to quantitatively predict tumor control during combined therapy for up to three weeks. The patterns depicted in **Figure 3b** shed light on the complexity arising from the interplay of various physical and biological processes. To facilitate a better understanding of our approach, three critical points should be emphasized. Firstly, the transformer model focuses on the mean output of each group (26 in total) to identify differences in temporal response, neglecting inter-animal variation. Secondly, the tumor volume change at a specific time point is interconnected with all preceding inputs. In other words, the cumulative effect is implicitly considered by the transformer model. Thirdly, our current goal is to identify the general trend in tumor control and potential synergy only in a qualitative manner. Quantitative analysis, if possible, needs to be postponed until additional biological correlates become available. Prior to delving into more in-depth discussion, we also would like to acknowledge that the results presented in this manuscript represent only the first step in the exploration of the PULSAR effect, and the interpretations drawn from the attention maps are largely speculative at this stage. The attention maps in our model, reflecting the link between two treatment sequences, are more abstract compared to the semantics in neural translation. More comprehensive investigations are need to corroborate definitive conclusions.

In the next phase of our experimental design, several limitations can be addressed to help us further improve the performance of the transformer model, as well as enhance the interpretability. First, the limited number of groups does not provide sufficient data diversity. How the data augmentation step impacts the model's performances awaits further investigation. Second, the prevalence of numerous zeros in the output sequence arises from volumes not measured on multiple days. A more continuous measurement of tumor volume would be advantageous. The preferred choice is to use the daily tumor volume as the output instead of the tumor volume change, as indicated in **Eq. 1**. Third, more effective embedding of the input sequence, currently represented by one bit for either radiation or anti-PD-L1 and four bits for positional embedding, could improve the predictive capability when more data are available. Fourth, this manuscript solely utilized tumor volume as output, lacking biological correlatives such as data on immune cell infiltration and their activities. Finally, it is also crucial for us to validate the model using different cell types and tumor models, as each is expected to demonstrate unique synergies.

Ordinary Differential Equation (ODE)-based mathematical models have been extensively used for analyzing complex biological systems, such as utilizing discrete-time equations to model tumor control under radiation and the interplay between immunotherapy and radiotherapy[23, 24]. In the future, we plan to adopt a similar approach and compare the performance between the ODE-based and AI approaches. The two may complement each other and enhance their interpretation. In our view, the transformer model offers several distinct advantages for investigating the PULSAR effect. For instance, predicting the tumor volume trajectory is a curve fitting task. With a limited number of parameters and compartment models, an ODE-based approach may not be able to achieve good fitting results, given the complexity of the biological processes involved in combined treatment. Moreover, other AI models such as recurrent neural network (RNN), are capable of identifying the underlying correlation in sequential signals. As a result, RNN would be a suitable candidate for predicting volume change, achieving comparable accuracy as the transformer model. However, RNN is not capable of providing information about the temporal interaction between two inputs. Apart from that, one distinct advantage of the transformer is that the arithmetic operation in attention maps between two groups helps pinpoint the respective contribution from each treatment as a function of days. As mentioned above, when a more continuous measurement of tumor volume or biological markers is available, we expect the cross-attention maps like Figure 3b to become much smoother and easier to interpret. An even broader question arises regarding the level of personalization in combined PULSAR and immunotherapy. Eventually, leveraging on the temporal interactions at different time points, we may identify optimum treatment strategy for each animal through reinforcement learning like previous studies[25, 26].

Whatever modeling tools selected, we posit that a deeper understanding of synergy in combined therapy can be gained through the examination of the PULSAR effect, an immunomodulatory impact that can be either inhibitory or stimulatory. The PULSAR effect can serve as a metric to allow us to assess whether the combined effect is greater than the sum of individual treatments. From a mechanistic perspective, it gauges the binding between PD-1 and PD-L1 in our study, contingent on multiple factors such as timing, dose, tumor microenvironment, and T cell dynamics. Its practical application is two-fold. On one hand, it can be used as an in-silico tool to explore new combinations that can maximize synergy and therapeutic efficacy, out of a substantial number of permutations. This may help interpret result discrepancy in published studies examining the synergy between radiation therapy (either daily fractionation or PULSAR) and immunotherapy[1-10]. As a result, a PULSAR trial including concomitant ICI blockade can be designed and performed more cost-effectively.

On the other hand, it can be combined with biomarker identification to offer us a completely new perspective on examining the interaction between radiation and ICI therapy at varying time points, through attention plots in **Figure 3b**. Leveraging on the advantages of AI modeling and in-silico studies (e.g., virtual simulator, smart trials), we might be able to delve into a number of causal relationships relevant to the PULSAR effect as exemplified below. We plan to collect information on two aspects: 1) the attention maps capable of identifying the influence of each treatment, and 2) the experimental measurement of biomarkers. When overlaying them along the same time axis, we will be able to better understand causal relationships and answer several relevant questions. For example, whether the PULSAR effect is instantaneous or takes time to build up? How the first radiation pulse recruits CD8+ T cells as a function of time? Whether the second radiation pulse kills both tumor cells and T cells, "resetting" the system to a new state? How each treatment alters the number and function of CD+8 T cells, $T_{reg}$ cells and others lymphocytes? What are the peripheral blood lymphocyte subsets at the time point when we see the maximum PULSAR effect? To be more specific, we envision the proposed framework may play a unique role in the following two aspects.

The first aspect is pertinent to timing and dose. If the interval between two pulses is too brief, the second radiation may eliminate newly recruited CD8+ T cells entering the tumor microenvironment. Conversely, excessively long intervals may hinder the optimal realization of therapeutic benefits from each treatment. In other words, it is essential to ensure that the arrival of new CD8+ T cells does not coincide with the timing of radiation pulses. Additionally, overly close intervals between the two pulses may also not be ideal, potentially stimulating tumor growth. For instance, a study using similar preclinical models demonstrated accelerated tumor growth and decreased survival when 3 Gy doses were administered daily[27, 28]. Likewise, the timing of administering anti-PD-L1 needs to be optimized, given the time-dependence of the PULSAR effect. It should be noted that optimizing timing will rely on tumor microenvironment and tumor type. In one previous study[11], the "cold" LLC tumor model exhibited the maximum synergy when radiation pulses were spaced 10 days apart, concurrently with the administration of the second anti-PD-L1 dose. By contrast, "hot" tumor models like colon carcinoma (MC38 cells) exhibited the maximum synergy immediately after the first radiation dose due to pre-existing immunity[11, 29].

The second aspect is pertinent to the temporal dynamics of biomarkers, particular T cells. There is a need for an investigation into how T cell infiltration is temporally influenced by radiation pulses concerning dose and spacing. Moreover, one study reported that T cells (mostly $T_{reg}$ cells) residing in the tumor during radiation display increased resistance to radiation, playing a critical role in overall tumor control[20]. By contrast, T cells in other body compartments exhibit less resistance to radiation. Although the authors did not administer a second dose of radiation after the initial one, it raises the possibility that a subpopulation of newly recruited T cells may infiltrate and develop into intratumoral T cells of different radiosensitivity. If that is the case, maximizing the PULSAR effect may demand a boost of either radiation dose or anti-PD-L1 dose when repetitive radiation is applied. Among other immune cells, we speculate that the PULSAR effect is most closely tied to $T_{reg}$ cells (i.e., the ratio between $T_{reg}$ cells and CD8+ T cells). In-vitro studies have demonstrated the combined effect of $T_{reg}$ regulation and radiation therapy[17-19]. In a study involving mice irradiated with 10 Gy to the right leg (prostate C1 cells)[19], $T_{reg}$ cells significantly increased in the spleen, lymph nodes, blood, and lung within two days after exposure, returning to normal levels by 10 days. The authors also noted three dose-dependent patterns: 1) $T_{reg}$ cells increased even after a low dose of 2 Gy, 2) mice receiving over 10 Gy irradiation exhibited double the fraction of splenic $T_{reg}$ cells compared to the control group, with no significant difference between 10 Gy and 20 Gy, and 3) fractionating a larger dose into smaller daily fractions (e.g., 10 Gy vs. 5×2 Gy, 20 Gy vs. 3×8 Gy) slightly reduced the production of $T_{reg}$ cells. Similar design and analysis will be included in the next phase of our study. Conjoining the temporal behavior of these biomarkers with the AI modeling will further validate the interpretation of the attention mechanisms and deepen our understanding in a number of causal relationships relevant to the PULSAR effect.

**Conclusion**

We developed a transformer based AI model to study the temporal synergy between radiation therapy and ICI therapy, aiming to answer whether synergy exists or if the efficacy resides within each modality. The model has potential to enhance our understanding of how the PULSAR effect depends on time, dose and T cell dynamics. It can also be used for in-silico modeling, facilitating the exploration of innovative treatment combinations to maximize synergy and therapeutic outcomes. Our work lays the foundation for exploring the PULSAR effect, however at this point several interpretations are largely speculative and further research with more comprehensive data will be needed. Advanced modeling techniques, combined with biomarker identification, will contribute to a more profound understanding of the biological mechanisms underlying the PULSAR effect.

**Data availability**

Data that support the findings of this study are available from the corresponding author upon request.

**Competing interests**

The author declares no competing interests.

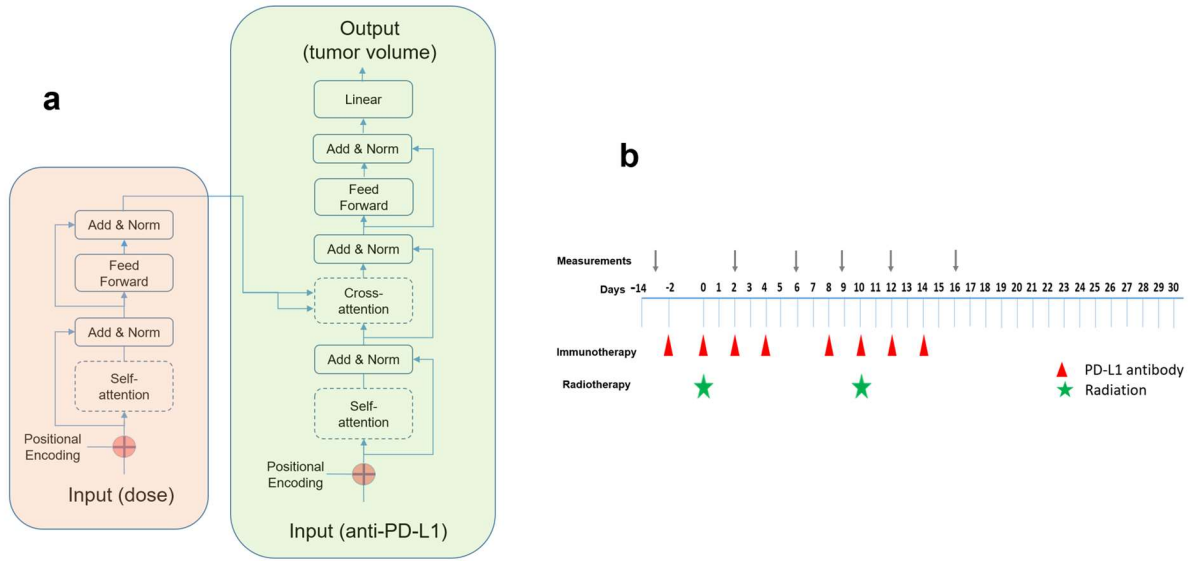

**Figure 1**. **a**. The transformer model comprising an encoder (left box) and a decoder (right box). For a given external stimulus at a specific time, the attention score indicates the strength of its interaction to those inputs in the past. An attention score (also known as alignment scores, based on the query and key vectors) would stay between zero to one, and multiplying it with the value vector yields the context vector to be connected to the feedforward block for predicting tumor volume change. **b.** The timing diagram of the combined therapy in our study. For immunotherapy, either anti-PD-L1 or isotype control was administered. For radiation, different doses were delivered (10, 15, 20, 40 Gy). The first pulse of radiation was always delivered 14 days after the implantation. Tumor volume measurements were carried out sequentially on certain days.

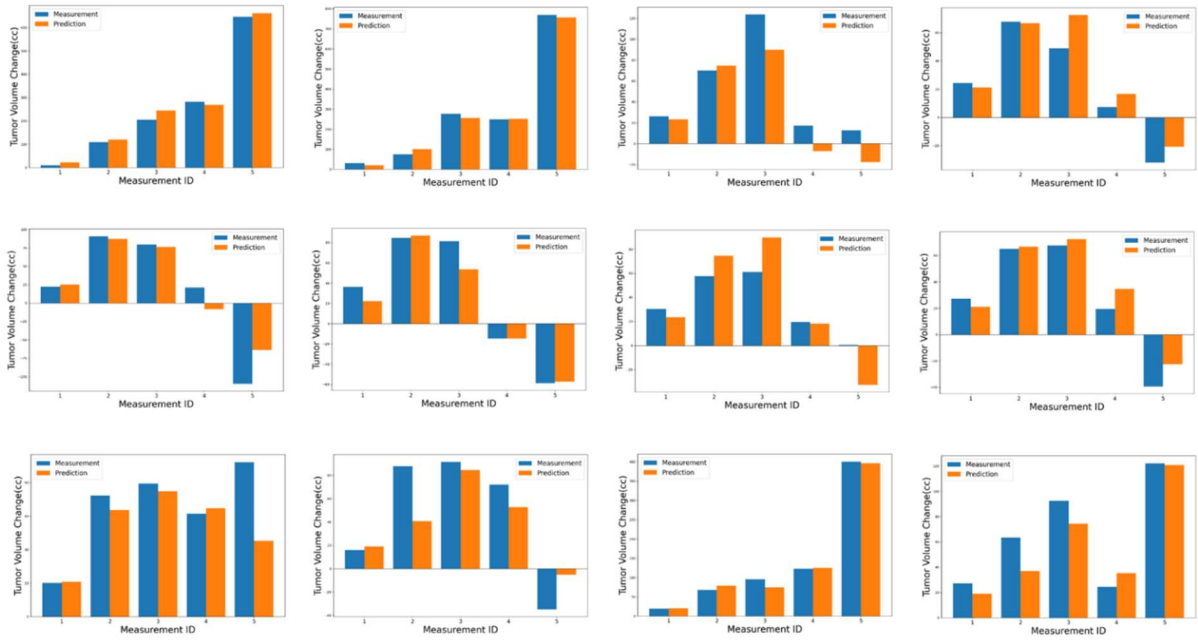

**Figure 2.** Predicted tumor volume change for groups 1-12 (top row: 1-4, second row: 5-8, third row: 9-12), in comparison with the results of measurements. Refer to Table S1 in supporting materials for the detailed schedule of each group.

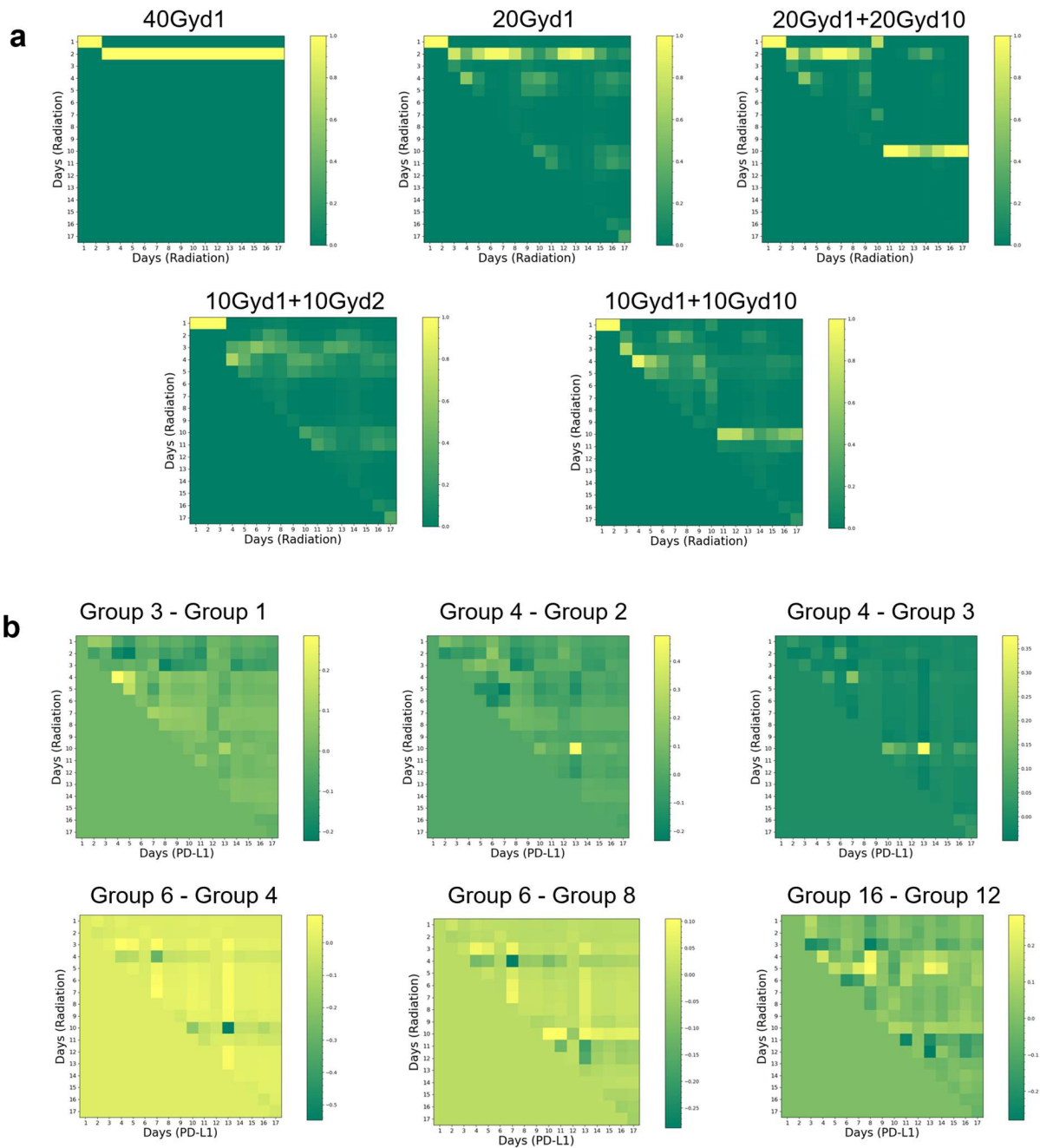

**Figure 3.** Visual representation of self-attention maps (**a**) and cross-attention maps (**b**) for selected groups. In self-attention maps, both horizontal and vertical axes represent the radiation sequence. As for cross-attention maps, the horizontal axis represents the anti-PD-L1 sequence, while the vertical axis represents the radiation sequence. The triangular pattern emerges from the application of a causal mask, which ensures that each location only has access to the locations that come before it.